\documentclass[epj,final]{svjour}
\usepackage{graphicx}
\usepackage{epsfig}
\usepackage{color}
\usepackage{amsmath,amssymb}

\newcommand{\PGR}[1]{{\color{black}#1}}

\renewcommand{\vec}[1]{{\mathbf{#1}}}

\begin{document}

\authorrunning{P.-G. Reinhard}
\title{Nuclear density-functional theory and fission of super-heavy elements}
\author{P.--G. Reinhard\inst{1}}
\institute{ 
Institut f{\"u}r Theoretische Physik, Universit{\"a}t Erlangen,
Staudtstrasse 7, D-91058 Erlangen, Germany
}
\date{\today / Received: date / Revised version: date}
\abstract{ 
We review the prediction of fission properties of super-heavy elements (SHE)
by self-consistent mean-field models thereby concentrating on the
widely used Skyrme-Hartree-Fock (SHF) approach. We explain briefly the
theoretical tools: the SHF model, the calibration of model parameters
together with statistical analysis of uncertainties and correlations,
and the involved computation of fission lifetimes. We present an
overview of fission stability in comparison to other decay channels
over the whole landscape of SHE reaching deep into the $r$-process
domain. The main emphasis lies on a detailed discussion of the
various ingredients determining eventually the fission properties.
The main result is that fission is an involved process which explores
many different influences with almost equal share, basic bulk
properties (also known as liquid-drop model parameters), pairing
strengths, and shell effects.
} 

\maketitle

\section{Introduction}

Since mid last century, it became possible to produce new elements not
naturally appearing on earth.  This triggered intense research on
transactinides \cite{Hyd57aR} proceeding with time to ever
heavier nuclei, coined super-heavy elements (SHE)
\cite{Nix72aR,Hof00aER}. The search is still going on and more new
elements are added every year to the list of SHE, see e.g.
\cite{HofS01,Oga04,Oga06,Greg06,Dvo08}.  Knowledge of exotic nuclei
beyond the valley of stability is crucial in understanding
astro-physical reactions and amongst them properties of SHE are
important in order to determine the upper end of the nucleosynthesis
flow \cite{Martinez-Pinedo.Mocelj.ea:2007,Arn07,Sto07aR}.  No surprise
then that SHE constitute an extremely active field of experimental and
theoretical research.  The Frankfurt theory
group under W. Greiner joined these efforts from the theoretical side
rather early \cite{Mos68a,Mos69a} and continued since. The theoretical
methods were much developing over the years. Today's standard, in
nuclear structure generally and also for SHE, are self-consistent
mean-field models in terms of nuclear density functional theory (DFT)
which came up in the 1970ies and have meanwhile reached high
descriptive power for nuclear structure and dynamics, for reviews see
e.g. ~\cite{Sto07aR,Rin96aR,Ben03aR,Vre05aR,Erl11a}.

The key question in the study of SHE is their stability, particularly
against spontaneous fission.  First estimates of stability were
obtained from analyzing shell structure.  The ultimate test, of
course, is a direct analysis of fission barriers and lifetimes. The
basic mechanism of nuclear fission as a many-body tunneling process
proceeding along steadily changing nuclear shapes has been understood
in terms of phenomenological shell models since long
\cite{Nix72aR,Bra72aR}.  However, self-consistent calculations of
fission lifetimes are extremely demanding and thus have come up only
recently, see e.g. \cite{Ber01a,War06,Sta09a} for calculations with
approximate collective masses or \cite{Sch09a,Erl12b} for fully
self-consistent calculations and for a recent review \cite{Bar15aR}.
It is the aim of this contribution to discuss predictions from
self-consistent models on fission barriers and lifetimes with a quick
glance at the competing reactions $\alpha$- and $\beta$-decay and
neutron emission.  Within the self-consistent models, we
concentrate in particular on the Skyrme-Hartree-Fock approach. We
discuss in detail its predictive power using methods of statistical
analysis related to the phenomenological calibration of the model.

The paper is outlined as follows: In section \ref{sec:model}, we
summarize the Skyrme-Hartree-Fock (SHF) model, the computation of
fission lifetimes, and explain statistical analysis for estimating
extrapolation uncertainties and parameter correlations. In section
\ref{sec:result}, we present and discuss the results starting from a
quantitative overview of various decay channels followed by an
extensive discussion of error estimates and correlations.

\section{Formal framework}
\label{sec:model}

\subsection{The Skyrme mean-field model}

We consider here nuclear DFT using the SHF energy-density functional
$\mathcal{E}_\mathrm{Sk}(\rho,\tau,{\cal J},{\bf j},{\bf\sigma})$,
which is expressed in terms of a few local densities and currents
obtained as sums over single-particle wave functions: density $\rho$,
kinetic density $\tau$, spin-orbit density ${\cal J}$, current ${\bf
j}$, spin density ${\bf\sigma}$, 
and pairing density $\xi$. This reads, e.g., for the density
\begin{equation}
  \rho_q(\mathbf{r})
  = 
  \sum_{\alpha\in q} f_\alpha v_{\alpha}^2|\varphi_{\alpha}(\mathbf{r})|^2
\end{equation}
where $q\in\{p,n\}$ labels proton or neutron density,
$\varphi_{\alpha}$ are s.p. wavefunctions, $v_{\alpha}^2$ the
corresponding pairing weights, and $f_\alpha$ pairing phase-space
factors (see below). It is convenient to formulate the functional in
terms of isoscalar density $\rho_0=\rho_n+\rho_p$ and isovector density
$\rho_1=\rho_n-\rho_p$. The same holds for the other densities and
currents.

Using these densities, the total energy is composed as
\begin{subequations}
\begin{equation}
E
= E_\mathrm{kin} 
  + \int \! d^3 r \; {\cal E}_\mathrm{Sk}
  + E_\mathrm{Coul}
  + E_\mathrm{pair}
  - E_\mathrm{corr}
  \quad,
\label{eq:Etot}
\end{equation}
\begin{eqnarray}
  E_\mathrm{kin} 
  &=&
  \int d^3r  \left(
      \frac{\hbar^2}{2m_p} \tau_p+  \frac{\hbar^2}{2m_n} \tau_n
   \right)
  \quad.
\label{eq:ekin}
\\
  {\cal E}_{\rm Sk}
  & = & 
        C_T^{\rho}\, \rho_{T}^{2}
      + C_T^{\rho,\alpha}\, \rho_{T}^{2}\rho_0^\alpha
      + C_T^{\Delta \rho} \, \rho_{T} \Delta \rho_{T}
\nonumber\\
      &&
      + C_T^{\tau} \, \rho_{T} \tau_{T}
      + C_T^{\nabla J} \rho_{T} \, \nabla\!\cdot\!\vec{J}_{T}
      + C_T^{J} \vec{J}^2_{T} 
     \;,
\label{eq:Eskeven}
\\
  E_\mathrm{Coul}
  &=&
  \frac{e^2}{2}\int d^3r\,d^3r'
   \frac{\rho_p(\vec{r})\rho_p(\vec{r}')}{|\vec{r}-\vec{r}'|}
\nonumber\\
   &&
   +\frac{3e^2}{4}\left(\frac{3}{\pi}\right)^{1/3}
   \int d^3r[\rho_p(\vec{r})]^{4/3}
   \quad,
\end{eqnarray}
\end{subequations}
where $e^2=1.44\,\mathrm{MeV}\,\mathrm{fm}$.  The ${\cal E}_{\rm Sk}$
here shows, for simplicity, only the part containing time-even
densities, because this defines already the basic model parameters
$C_T^\mathrm{(typ)}$ and only this part counts in stationary
calculations.
The tensor spin-orbit term $\propto C_T^{J}$ is, in fact, more
involved \cite{Ben07a}. But its details have little impact on basic
bulk properties.  Thus it is usually ignored at all with setting
$C_T^J=0$.
For a detailed discussion of the functional, its parameters, time-odd
terms and other options see \cite{Erl11a}.

The correlation energy $E_\mathrm{corr}$ may contain several
contributions.  The least to do is the correction for the spurious
center-of-mass energy
$E_\mathrm{cm}=\frac{1}{2mA}\langle\hat{\vec{P}}_\mathrm{cm}^2\rangle$
where $mA$ is the total mass and
$\hat{\vec{P}}_\mathrm{cm}=\sum_{n=1}^A\hat{\vec{p}}_n$ the total
momentum. The $\hat{\vec{P}}_\mathrm{cm}^2$ is, in fact, a two-body
operator. Some Skyrme parametrizations simplify that by using only its
diagonal term of, for
details see \cite{Ben03aR,Erl11a}. Deformed nuclei require also a
correction for rotational projection and soft surface vibrations
\cite{Rei75a,Klu08a}. These will be discussed in connection with the
fission path in section \ref{sec:CHF}.

The pairing functional reads
\begin{subequations}
\begin{eqnarray}
  \mathcal{E}_\mathrm{pair}
  &=&
  \frac{1}{4} \sum_{q\in\{p,n\}}V_\mathrm{pair,q}
  \int d^3r \xi^2_q
  \left[1 -\frac{\rho}{\rho_{0,\mathrm{pair}}}\right]
  \quad.
\label{eq:ep-op2}
\end{eqnarray}
Letting $\rho_{0,\mathrm{pair}}\rightarrow\infty$ suppresses
the density-dependent part and recovers what is called volume pairing.
Surface pairing is obtained by setting $\rho_{0,\mathrm{pair}}=0.16$
fm$^{-3}$. Most generally, one allows $\rho_{0,\mathrm{pair}}$ to be a
free parameter. Optimizing the pairing functional this way, one obtains
as optimum a mix of volume and surface pairing \cite{Klu09a}.
The pairing functional (\ref{eq:ep-op2}) is complemented by a cutoff in
s.p. space \cite{Rin80aB}.  We use a soft cutoff factor
\begin{equation}
  f_\alpha
  =
  \left[1+
    \exp{\left((\varepsilon_\alpha-(\epsilon_F+\epsilon_{\rm cut}))
            /\Delta\epsilon\right)}
  \right]^{-1}
  \quad,
\label{eq:softcut}
\end{equation}
where $\Delta\epsilon=\epsilon_{\rm cut}/10$,
$\epsilon_\alpha$ is the single particle energy of the state
$\alpha$ and $\epsilon_F$ is the chemical potential.  The original
recipe used a fixed cutoff energy $\epsilon_{\rm cut}=$ 5-10 MeV
\cite{Bon85a,Kri90a}. Here, we determine $\epsilon_{\rm cut}$ such the
the active s.p. space has a fixed size $\propto N^{2/3}$, yielding a
cutoff which depends on the actual nucleon number $N_q$ \cite{Ben00c}
\begin{equation}
  \epsilon_{\rm cut}
  \quad\longleftrightarrow\quad
  \sum_\alpha f_\alpha
  =
  N_q+\eta_\mathrm{cut}N_q^{2/3}
  \quad.
\label{eq:cutparam}
\end{equation}
\end{subequations}
We use $\eta_\mathrm{cut}=1.65$ in the following calculations.

The mean-field equations are derived variationally from the given
energy functional.
Pairing is performed at the level of the BCS approximation.

The SHF functional sets only a framework. The model parameters
$C_T^\mathrm{(typ)}$ remain yet to be determined.  In the early days
of nuclear DFT, the quality of nuclear ab-initio calculations was
poor. Thus the parameters of the functional had been adjusted
empirically. The most systematic way to do this are least-squares
($\chi^2$) fits \cite{Bev69aB} used for nuclear DFT first in
\cite{Fri86a} and being meanwhile a standard technique for calibration
of self-consistent mean-field models. One finds in the literature a
great variety of functional development using $\chi^2$-fits to
empirical data. For example, some concentrate on spherical nuclei with
negligible correlation effects \cite{Klu09a}, others are particularly
concerned with deformed nuclei \cite{Nik04a,Nik08a,Kor12a}, still
others try to adjust also spectra of s.p. energies
\cite{Bro98a,Kor14a}. Although differences occur in details due to
different choice of fit data, it is reassuring that all these fits
provide similar and high-quality results for the reproduction of
global nuclear properties.  Nuclear many-body theory has made in the
meantime enormous progress, see e.g. \cite{Hag08a,Nav09a,Gar16a}.
Nonetheless, the precision of these ab-initio models does not yet
suffice for a high-quality description of nuclear properties.  Thus it
is still state of the art to adjust the DFT parameters empirically.
We will address that briefly in section \ref{sec:fit}.

The parameters of the SHF functional (\ref{eq:Eskeven}) are initially
the $C_T^\mathrm{(typ)}$. Some of them can be associated with an
immediate meaning. For example, the $C_T^{\Delta \rho}$ characterize
surface tension and $C_T^{\nabla J}$ the spin-orbit strength. The
volume parameters, however, are less transparent. It is advantageous
to express them through the nuclear matter parameters (NMP), i.e. the
basic properties of symmetric matter: the ground state properties
equilibrium binding energy $E/A_\mathrm{eq}$ and equilibrium density
$\rho_\mathrm{eq}$, the static response properties incompressibility
$K$, symmetry energy $J$ and density dependence of symmetry energy
$L$, and the dynamic response properties (isoscalar) effective mass
$m^*/m$, and Thomas-Reiche-Kuhn sum-rule enhancement
$\kappa_\mathrm{TRK}$ which characterizes isovector effective mass,
for a detailed definition see e.g. \cite{Erl11a}. There is a
one-to-one mapping between the SHF volume parameters and NMP
\begin{equation}
 C_{0,1}^{\rho}, C_{0,1}^{\rho,\alpha}, \alpha, C_{0,1}^{\tau}
 \Leftrightarrow
 \frac{E}{A}\Big|_\mathrm{eq}, \rho_\mathrm{eq}, K, J, L, \frac{m^*}{m}, \kappa_\mathrm{TRK}
\end{equation}
such that both ways of handling the model parameters are equivalent. 
We will use later on the form in terms of NMP.

\subsection{Variety of SHF parametrizations}
\label{sec:params}

\begin{figure}
\centerline{\includegraphics[width=\linewidth]{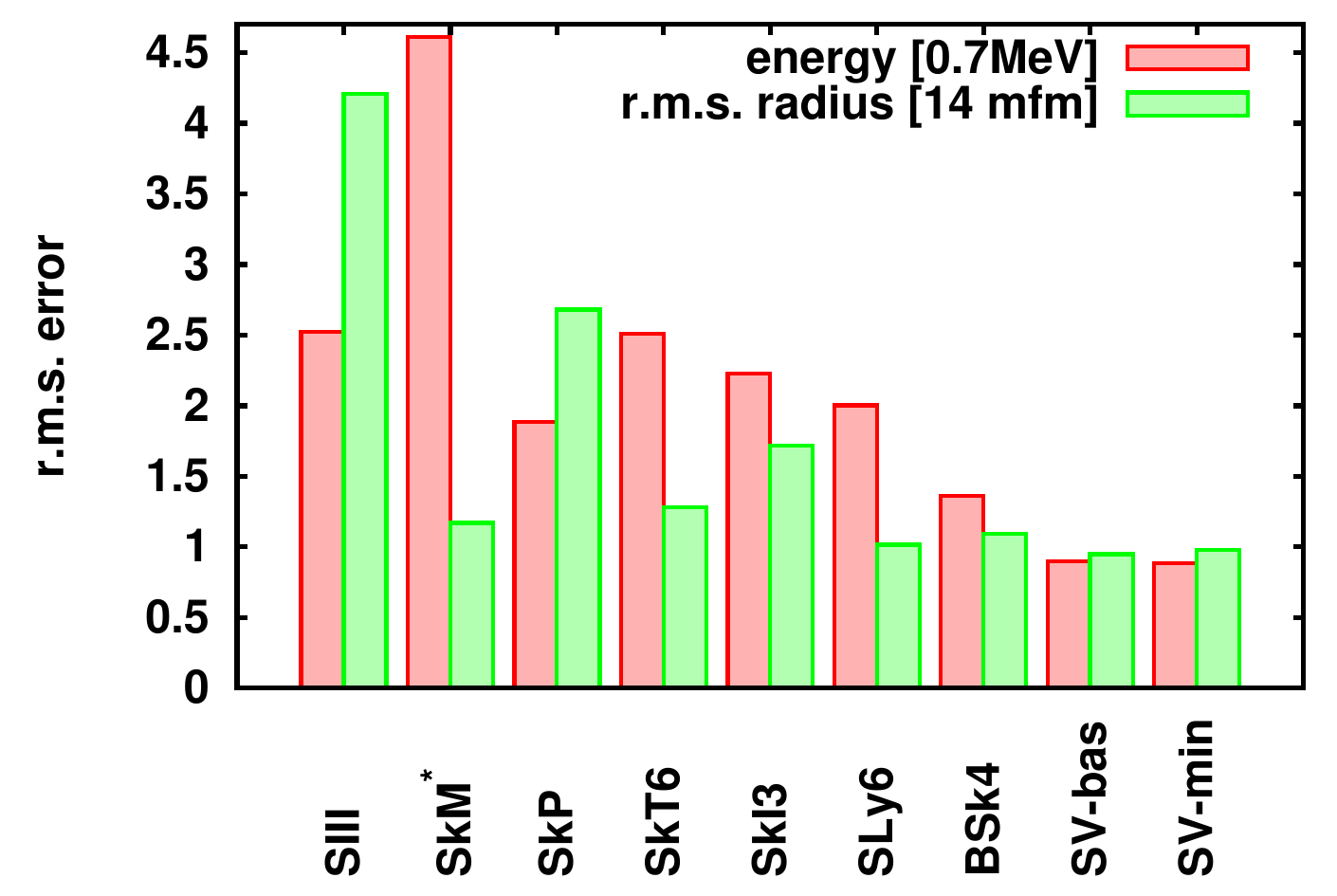}}
\caption{\label{quality-forces}
Quality of a couple of Skyrme parametrizations in terms of
r.m.s. error on binding energy and charge r.m.s. radius.
The parametrizations are sorted along time of publication:
SIII \cite{Bei75a}, SkM$^*$ \cite{Bar82a}, SkP \cite{Dob84a}, SkT6
\cite{Ton84a}, SkI3 \cite{Rei95a}, SLy6 \cite{Cha98a},
BSk4 \cite{Gor03a}, SV-bas and SV-min \cite{Klu09a}.
The average errors are computed over the large pool of fit data from
\cite{Klu09a} which contains semi-magic nuclei over a wide range of
sizes and isotopes.
  }
\end{figure}
Empirical adjustment leaves some freedom in the choice of reference
data and relative weights of them. Thus there exists in the literature
a variety of parametrizations.  Figure \ref{quality-forces} compares a
selection of typical and widely used para\-me\-trizations as they have
developed in the course of time.  Means of comparison is the quality
of binding energy and charge r.m.s. radii in terms of the
r.m.s. deviation to experimental data for the selection of semi-magic
nuclei of \cite{Klu09a}.  The figure shows how parametrizations have
improved over the decades. This is due to gathering more experience
with systematic fits over the years and, more important, to the
appearance of new data on exotic nuclei which help enormously to pin
down the properties of a functional. A comment is in order about the
two earliest parametrizations. The first one, SIII, concentrated on
energies with the consequence that these are already quite well
reproduce while the radii are a bit off. The next move, SkM$^*$,
worked on a correct reproduction of excitation properties and fission
barriers. This involves a large improvement on r.m.s. radii with
sacrifices at the side of energies. Nonetheless, SkM$^*$ served for a
long period almost as standard because it provides a well equilibrated
reproduction of a large set of observables. From then on, improvements
did proceed steadily and slowly \PGR{and the process is still going
  on. In the meantime came up parametrizations which come down to an
  average error on energies of 0.5 MeV and on r.m.s. radii on 0.012 fm
  \cite{Gor13a}. However, this often requires corrective terms not
  included in the present SHF model and the gain is subtle. Thus we
  confine the present study to the above list of parametrizations.  }
After all, we emphasize that most parametrizations deliver a
comparable and good description of ground-state data.
\begin{table}
\PGR{
\begin{center}
\begin{tabular}{|l|rr|rrr|}
\hline
 & \multicolumn{1}{|c}{$K$}
 & \multicolumn{1}{c}{$m^*/m$}
 & \multicolumn{1}{c}{$J$}
 & \multicolumn{1}{c}{$L$}
 & \multicolumn{1}{c|}{$\kappa_\mathrm{TRK}$}
\\
\hline
Gogny D1 &  229 & 0.67 & 31 & 18 & 
\\
\hline
SkI3 & 258 &   0.58 &    35 &   212 &   0.25
\\
SLy6 & 230 &   0.69 &    32 &   100 &   0.25
\\
SkT6 & 236 &   1.00 &    30 &    64 &   0.00
\\
SkM$^*$ & 217 &   0.79 &    30 &    95 &   0.53
\\
SkP & 201 &   1.00 &    30 &    41 &   0.35
\\
BSk4 &  237 &  0.92 &    28 &    27 &   0.18
\\
SV-bas & 234 &   0.90 &    30  &  68 &   0.40
\\
SV-min & 222 &   0.95 &    31 &    93 &   0.08
\\
UNEDF2 & 240 &   0.93 &    29 &    40 &   0.25
\\
\hline
NL-Z2 & 175 & 0.58 & 39 & 126 & 0.72
\\
DD-PC & 185 & 0.57 & 35 & 82 & 0.75
\\  
DD-ME &  250 & 0.56 & 32 & 6 & 0.79
\\
\hline
\end{tabular}
\end{center}
\caption{\label{tab:NMP} Nuclear matter parameters (NMP)
  for the for the various parametrizations
  used in this paper. The $K$, $J$, and $L$ are given in MeV;
  $m^*/m$ and $\kappa_\mathrm{TRK}$ is dimensionless.
  In case of RMF (last three entries), $m^*/m$ and
  $\kappa_\mathrm{TRK}$ stand for the values at momentum $k=0$.
  }
}
\end{table}
They differ, however, in other \PGR{
features. Table
  \ref{tab:NMP} exemplifies this for NMP.  Similar variances of
  results are also seen for giant resonances, isotopic trends, and
  particularly fission barriers which will be discussed later.}
In addition to the above listed parametrizations, we will employ in
section \ref{sec:trend} a set with a systematic variation of model
parameters.

\subsection{Calibration and statistical analysis}
\label{sec:fit}

The principles of $\chi^2$-fits are simple.  The model under consideration,
here SHF, produces for given model parameters $\mathbf{p}$ a
great manifold of expectation values of observables $A=A(\mathbf{p})$. We select a
subgroup of fit observables $\hat{\mathcal{O}}_f$ for which we have
reliable experimental data and from which we assume that they can be
described reliably well by DFT.  The quality of the model is
quantified by
\begin{equation}
  \chi^2(\mathbf{p}) 
  =
  \sum_f
  \frac{(\mathcal{O}_f(\mathbf{p})-\mathcal{O}_f^\mathrm{(exp)})^2}
       {\Delta\mathcal{O}_f}
\end{equation}
where $f$ runs over the pool of fit data,
$\hat{\mathcal{O}}_f^\mathrm{(exp)}$ stands for the corresponding
experimental values, and $\Delta\mathcal{O}_f$ is an adopted error
which regulates the relative weight of different observables. It
accounts mainly for the expected theoretical reliability for this
observable and to lesser extend also to the experimental uncertainty.
There is an automatic feedback built in: The adopted errors
$\Delta\mathcal{O}_f$ are chosen correctly if the final $\chi^2$ is of
order of the number of data points (for a detailed discussion see
\cite{Bev69aB,Dob01b}). The best fit is obviously given by that
parametrization $\mathbf{p}_0$ which minimizes $\chi^2$. What thus
remains is to find the absolute minimum of the multi-dimensional
function $\chi^2(\mathbf{p})$. This is a non-trivial task for which,
however, a great deal of know-how is around.

\begin{figure}
\centerline{\includegraphics[width=\linewidth]{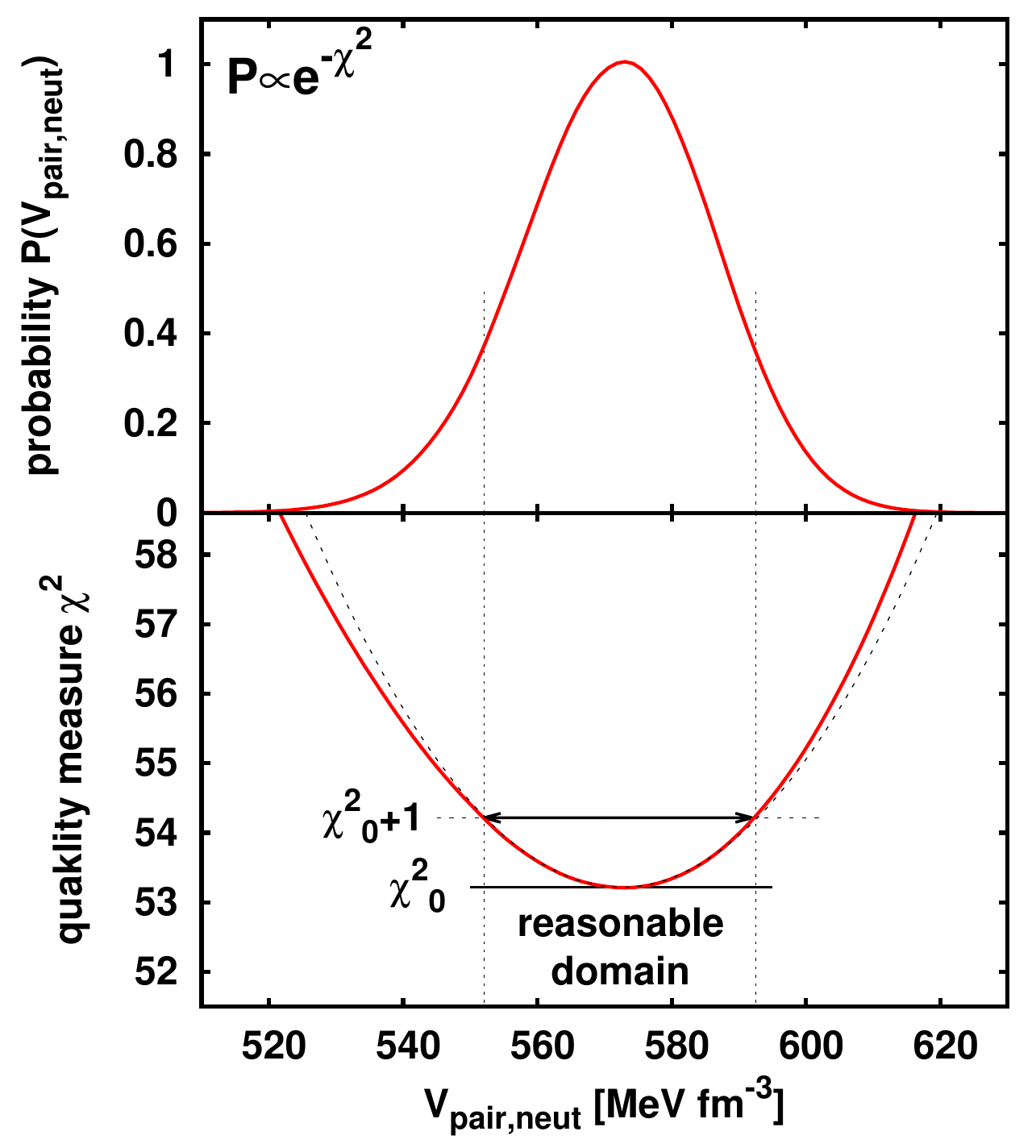}}
\caption{\label{fig:Reasonabledomain-vpairn}
Illustration of the statistical interpretation of $\chi^2$.
Lower panel: $\chi^2$ for the case of the parametrization
SV-min \cite{Klu09a} as function of  $V_\mathrm{pair,n}$; the
fine dashed line indicates the quadratic approximation around the
minimum.
Upper panel: corresponding probability $P$ of the parametrization.
}
\end{figure}
There is more in the $\chi^2$-scheme than just determining the minimum
$\chi^2_0=\chi^2(\mathbf{p}_0)$. The vicinity of the minimum carries also
worthwhile information. Parameters which lie near the
minimum can still provide a reasonable good description.  How good
depends, again, an the actual $\chi^2$ landscape.  This is illustrated
in figure \ref{fig:Reasonabledomain-vpairn} whose lower panel shows a
cut through the landscape along the pairing parameter
$V_\mathrm{pair,n}$.  It is obvious that $\chi^2$ remains
acceptably low in the immediate neighborhood of the minimum. This is
quantified in terms of the corresponding ``probability of a
parametrization'' $P(\mathbf{p})$ whose cut along $V_\mathrm{pair,n}$
is shown in the upper panel.  The $P(\mathbf{p})$ leads to the
statistical interpretation of $\chi^2$.  The maximum probability $P$
is found, of course, for minimal $\chi^2$ and the curvature of
$\chi^2$ around the best fit $\mathbf{p}_0$ determines the speed of
growth.  The range of reasonable parameters is defined as the area of
$\mathbf{p}$ for which $\chi^2\leq\chi^2(\mathbf{p}_0)+1$
\cite{Bev69aB}. The $\chi^2(\mathbf{p})$ near the minimum is close to
a parabolic function (compare full and dashed line in the lower panel
of figure \ref{fig:Reasonabledomain-vpairn}).  Thus the range of
reasonable parameters covers an ellipsoid in $\mathbf{p}$-space. The
probability distribution $P(\mathbf{p})$ allows to compute statistical
averages of any observable ${A}={A}(\mathbf{p})$.  For the average, we
have usually $\overline{A}=\int
d\mathbf{p}\,P(\mathbf{p})\,A(\mathbf{p}) \approx
A(\mathbf{p}_0)$. The uncertainty $\Delta A=\sqrt{\Delta^2A}$ of an
observable is obtained simply from
\begin{equation}
  \Delta^2A
  =
  \overline{(A-\overline{A})^2}
  =
  \int d\mathbf{p}\,P\;(A-\overline{A})^2
  \quad.
\end{equation}
It quantifies the extrapolation uncertainty propagated from the
uncertainties of the model parameters.  Still more information can be
gained from mixed uncertainties $\overline{\Delta A\Delta B}$ computed
in similar fashion.  They allow to deduce the amount of statistical
correlation, also called alignment between two observables. It is
defined as
\begin{equation}
  r_{AB}
  =
  \frac{\overline{\Delta A\Delta B}}
       {\sqrt{\overline{\Delta^2A}}\,\sqrt{\overline{\Delta^2B}}}
  \quad.
\end{equation}
A value $r_{AB}=\pm 1$ means that the two observables are fully
(anti-)correlated, i.e. knowledge of $B$ does not add any new
information to knowledge of $A$ (within the given model).
In contrast, a value $r_{AB}=0$ means that $A$ and $B$ are
uncorrelated, thus fully independent within the model. Often, one is
not interested on the sign of $r_{AB}$ and considers the
coefficient of determination (CoD) $r_{AB}^2$ \cite{Glantz}. This is
what we will present in the results later on.

Correlation analysis is compact and instructive, but often too
compact.  A visual impression of correlations can be visualized by
trend analysis. An example will be given in section \ref{sec:trend}.

\subsection{Constrained mean field and the collective path}
\label{sec:CHF}

Fission is an extension of collective quadrupole motion to finally two
fragments.  It proceeds through a succession of quadrupole deformed
mean fields, called the collective path. We generate the path by
imprinting a dedicated deformation using quadrupole-constrained
mean-field equations
\begin{subequations}
\begin{equation}
  \left(\hat{h}-\epsilon_{\rm F}\hat{N}-\lambda\hat{Q}_{20}\right)
  |\Phi_{\alpha_{20}}\rangle
  =
  {\cal E}|\Phi_{\alpha_{20}}\rangle
  \quad,
\label{eq:Cmfeq}
\end{equation}
where $\hat{h}$ is the SHF mean field Hamiltonian (depending on
the local densities), $\epsilon_{\rm F}$ is the Fermi energy, and
$\lambda$ the Lagrange parameter for the quadrupole constraint.
The $\hat{Q}_{20}$ is the quadrupole operator and $\alpha_{20}$ its
dimensionless expectation value, i.e.
\begin{eqnarray}
  \hat{Q}_{20}
  &=&
  r^2Y_{20}g_{\rm cut}({\bf r})
  \quad,
\\
  \alpha_{20}
  &=&
  \frac{4\pi}{5}
  \frac{\langle\Phi_{\alpha_{20}}|r^2Y_{20}|\Phi_{\alpha_{20}}\rangle}
       {Ar^2}
  \quad,
\label{eq:label}
\end{eqnarray}
\end{subequations}
with $A$ the total particle number and $r$ the r.m.s. radius.  Note
that we consider here only axially symmetric quadrupole deformations,
i.e. only the 0 component of the more general $\alpha_{2m}$.The
quadrupole operator is modified by a damping function $g_\mathrm{cut}$
which cuts the quadratic growth at large distances to avoid artifacts
from deep wells at the bounds of the numerical box \cite{Rut95a}.  The
equations are solved with an extra iterative loop to maintain a wanted
value of $\alpha_{20}$ \cite{Cus85a}. This is done for a dense set of
deformations $\alpha_{20}$ which yields the collective path as
a series of mean-field states $\{|\Phi_{\alpha_{20}}\rangle\}$ along
which the collective motion can evolve.

Having the path, we can compute a raw collective potential
$\mathcal{V}$ as as the SHF energy along the path states,
i.e. $\mathcal{V}(\alpha_{20})= E[\Phi_{\alpha_{20}}]$. The
self-consistent evaluation of the collective mass ${M}(\alpha_{20})$
for motion along $\alpha_{20}$, also called ATDHF mass, is rather
involved as it requires to compute the dynamical response to changing
shape, for details see \cite{Klu08a,Rei87aR}.  
In similar fashion, one computes also the momentum of
inertia $\Theta$ for rotation orthogonal to the nuclear axis.
With these collective
masses and widths,
$\mu_\mathrm{quad}(\alpha_{20})=\langle\Phi_{\alpha_{20}}|\hat{P}_{20}^2|\Phi_{\alpha_{20}}\rangle$ and
$\mu_\mathrm{rot}(\alpha_{20})=\langle\Phi_{\alpha_{20}}|\hat{J}_{x}^2+\hat{J}_{y}^2|\Phi_{\alpha_{20}}\rangle$, 
we compute the quantum correction to the potential yielding finally
the true collective potential as
\begin{eqnarray}
  V(\alpha_{20}) 
  &=& 
  \mathcal{V}(\alpha_{20}) 
  -
  \frac{\mu_\mathrm{quad}}{2M}
  - 
  \frac{1}{4\lambda}\partial_{\alpha_{20}}^2\mathcal{V}
  -
  \frac{\mu_\mathrm{rot}}{2\Theta}
  \quad.
\label{eq:Vcorr}
\end{eqnarray}
With $V(\alpha_{20})$, ${M}(\alpha_{20})$, and $\Theta(\alpha_{20})$,
we dispose of all ingredients for the collective Hamiltonian
which determines quadrupole vibrations and fission
\cite{Klu08a,Sch09a,Rei87aR}.

\begin{figure}
\centerline{\includegraphics[width=\linewidth]{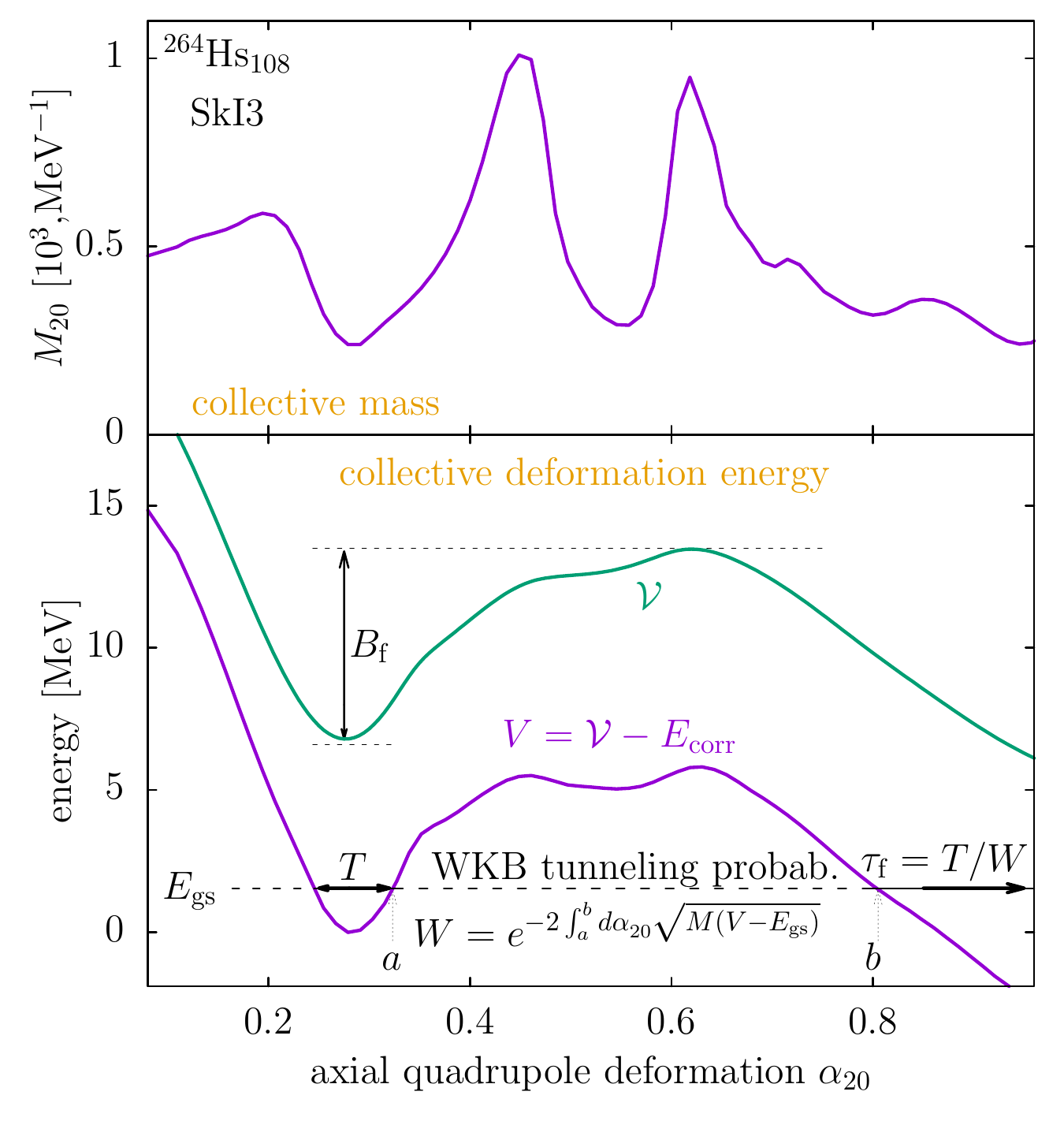}}
\caption{\label{PES-mass-264Hs} Potential energy surface (lower panel)
  and inverse collective mass (upper panel) along the fission path for
  $^{264}$Hs computed with the SHF parametrization SkI3
  \cite{Rei95a}. For the potential energy, one distinguishes the raw
  energy $\mathcal{V}$ and the fission potential $V$ including
  quantum correction, see eq. (\ref{eq:Vcorr}).
  The ground-state energy $E_\mathrm{gs}$ is indicated by a horizontal
  line, dashed in the classically forbidden (tunneling) regime and full in the
  classically  allowed regime. The limits of the tunneling
  regime  are indicated by $a$ and $b$ and $W$ is the tunneling
  probability. The $T$ stands for the oscillation time of the ground
  state within the binding pocket and $\tau_\mathrm{f}$ is the fission
  lifetime. 
}
\end{figure}

\PGR{ The above review of fission calculations is very sketchy. The
  treatment of pairing, in particular, covers more subtleties.  One
  point is that the BCS approximation to full HFB weakens if a
  considerable amount of continuum states becomes occupied. This may
  play a role for the weakly bound nuclei at the edges of the large
  scans of isotopes in the following. But the observed trends still
  remain relevant and BCS is valid for the well bound SHE which are
  discussed in quantitative detail. A more important point concerns
  controlling of particle.  A projection on
  exact particle number would be ideal. This, however, is inhibited
  for typical nuclear density functionals \cite{Dob07a}.  BCS provides
  at least conservation of particle number in the average by virtue of
  the particle-number constraint. Thus all states along the path
  represent the same average proton and neutron number.  To maintain
  this crucial feature for the coherent superposition of states in the
  collective ground state and along fission, we carry forth the
  particle-number constraint into the collective Schr\"odinger
  equation. This somewhat involved }
procedure is explained in great detail in \cite{Sch09a,Erl12b} and
reviewed briefly in figure \ref{PES-mass-264Hs}.  The basic structure
of a fission potential is already set by the raw potential
$\mathcal{V}$, namely a binding pocket at some small deformation
followed by a growth towards a maximum (barrier) then turning to a
steady decrease due to unhindered Coulomb repulsion.  An immediate
measure of stability is the fission barrier which we take here for
simplicity as
$B_\mathrm{f}=\mathcal{V}_\mathrm{max}-\mathcal{V}_\mathrm{min}$, the
difference between maximum and ground-state minimum in the raw
potential.  The quantum corrections change the potential curve at
a quantitative level, typically reducing the barrier by about 1-2 MeV
which, however, has sizable consequences for fission lifetimes
\cite{Rei75a,Rei87aR,Rei78a}. The collective mass $M$ has a much
fluctuating structure produced by level crossings near the Fermi
energy while the momentum of inertia $\Theta$ (not shown) is a smooth
function steadily increasing with deformation \cite{Rei84b}. Fission
lifetime is computed as tunneling dynamics along the collective
quadrupole momentum. First, one computes the ground state in the
pocket by solving the collective Schr\"odinger equation in the three
dimensions set by deformation $\alpha_{20}$ and two rotation angles
(about $x$ and $y$ axis).  This defines at fully quantum mechanical
level the ground state energy $E_\mathrm{gs}$. The repetition rate $T$
for the oscillations in the pocket are computed semiclassically by
$T=\hbar\int_a^bd\alpha_{20}^{\mbox{}}\sqrt{M/(V-E_\mathrm{gs})}$.
The tunneling probability $W$ is calculated in similar fashion, see
figure \ref{PES-mass-264Hs}. Both together yield finally the fission
lifetime $\tau_\mathrm{f}=T/W$.  It is to be noted that fission
dynamics is computed within three-dimensional quadrupole dynamics
related to axially symmetric configurations while full quadrupole
dynamics would explore a five-dimensional space covering also triaxial
shapes.  The limitation is justified because the fission process
follows a rather narrow fission valley. We checked that a
five-dimensional calculation of the ground state does not make much of
a difference in the regime SHE \cite{Sch09a,Erl12b}.

After all, we see that fission lifetime $\tau_\mathrm{f}$ is a highly
complex observable composed from many different ingredients.  Pocket
and barrier are generated by an interplay of Coulomb pressure and
shell effects \cite{Bra72aR}. The collective mass gathers influences
of all level crossings along the path. The collective ground state
energy determines the repetition rate $T$ and entry point $a$ as well
as exit point $b$ of the tunneling region whose width $b-a$ has large
impact on the tunneling probability $W$. Considering all these highly
sensitive influences, it is more than surprising that theory can
deliver at all a decent value for $\tau_\mathrm{f}$. However, we
should be prepared to see large differences in predictions with tiny
variations of the models parametrization.

\section{Results and discussion}
\label{sec:result}

\subsection{Systematics of barriers and lifetimes over the landscape
  of SHE}
\label{sec:syst}

\begin{figure*}
\centerline{\includegraphics[width=\linewidth]{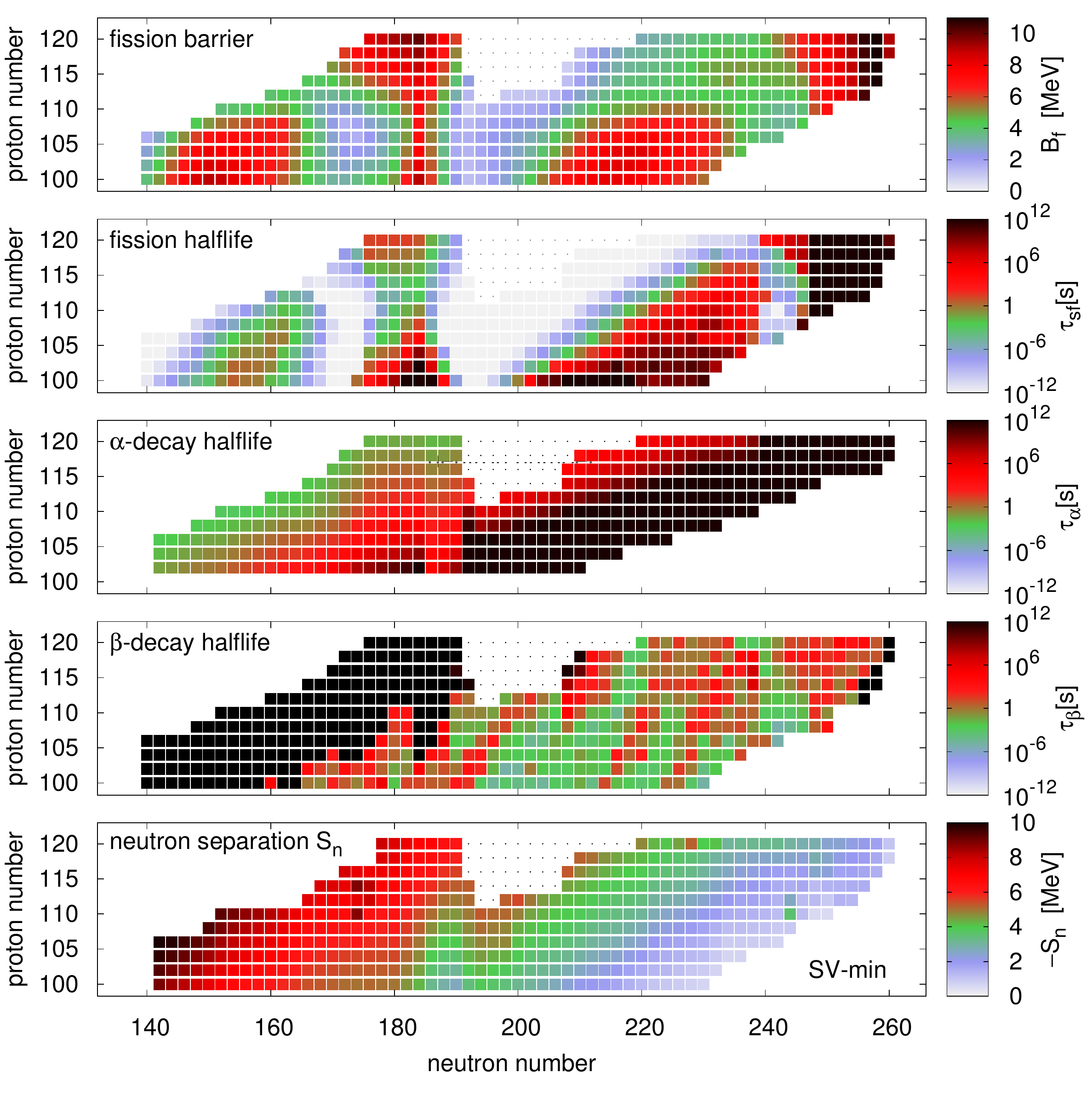}}
\caption{\label{fig:systematics} Systematics of fission barrier
  (upper), lifetimes for three different decay channels, and neutron
  separation energies (lower) over the
  landscape of SHE computed with SV-min \cite{Klu09a}. All cases use
  comparable color (grayscale) scheme. More stability is indicated by
  red (dark) and less instability by blue color (light gray).
  The region of triaxial fission, indicated by dots, has not been
  computed and is thus not considered here.}
\end{figure*}
We start the presentation with, so to say, the final result and we use
for that the SHF parametrization SV-min which promises to provide
reliable estimates of fission properties, as we will
work out later. Figure \ref{fig:systematics} shows the systematics of
quantities characterizing the stability of SHE: fission barriers,
fission lifetimes, $\alpha$-decay lifetimes, and neutron separation
energies.  The computation of fission barrier and lifetimes was
explained in section \ref{sec:CHF}. The $\alpha$- and $\beta$-decay lifetimes
$\tau_\alpha$ are computed using the Viola Seaborg-formula
\cite{Vio66a}.  The $\beta$-decay lifetimes $\tau_\beta$ were computed
in perturbation theory evaluating explicitly the $\beta$-transition
matrix elements to the final odd-odd nucleus, for details see
\cite{Erl12b}. The neutron separation energies are simply the energy
differences between the given nucleus and the neighboring odd nucleus
obtained by removing one neutron. For the thus necessary computation
of odd-odd and odd-even nuclei see \cite{Pot10a}.

The nuclear landscape covered in figures \ref{fig:systematics} reaches
deep into the region of neutron instability (see the panels neutron
separation energies and $\beta$ halflifes). Note that the neutron rich
region cannot be reached in a laboratory on earth but may play a role
at the upper end of the astro-physical $r$ process.  Let us look first
at fission barriers and lifetimes. At first glance, they show, of
course, the same trends. Large fission stability is seen at regions of
near shell closures (Z/N$\approx$108/150, N$\approx$184) and toward
extremely large neutron numbers, deep in the $r$ process region.
Between the islands of stability extend swamps of extremely short
fission lifetimes.  These quick fluctuations of fission lifetimes over
the landscape are due to shell effects.  Besides these shell
fluctuations, there is an interesting trend with proton number. With
increasing proton number $Z$, fission lifetimes become comparatively
shorter. The point is that barrier height $B_\mathrm{f}$ is not everything. The
width of the barrier has also large impact on the tunneling
probability $W$ and the widths shrink toward increasing $Z$. Even
though, we still encounter considerable fission stability at the upper
island Z/N$\approx$120/180.

The $\alpha$-decay lifetimes show smooth trends throughout and in the
average they have similar order of magnitude as fission
lifetimes. This yields an interesting interplay with fission
lifetimes. Near the islands of fission stability, $\alpha$-decay
prevails. But the corresponding $\alpha$-chains are interrupted as
soon as they cross a swamp of fission instability. This is a feature which
complicates experimental assessment of SHE.  The same interplay is
seen at the extreme neutron rich side.  Both lifetimes, fission and
$\alpha$-decay, become very large there. This may give a chance for
accumulation of SHE in stellar matter. We have to keep in mind,
however, that barriers and $Q_\alpha$ may change in the presence of
neutron gas \cite{Uma15a}.

%

\subsection{Barriers for various parametrizations}
\label{sec:barr}

\begin{figure}
\centerline{\includegraphics[width=\linewidth]{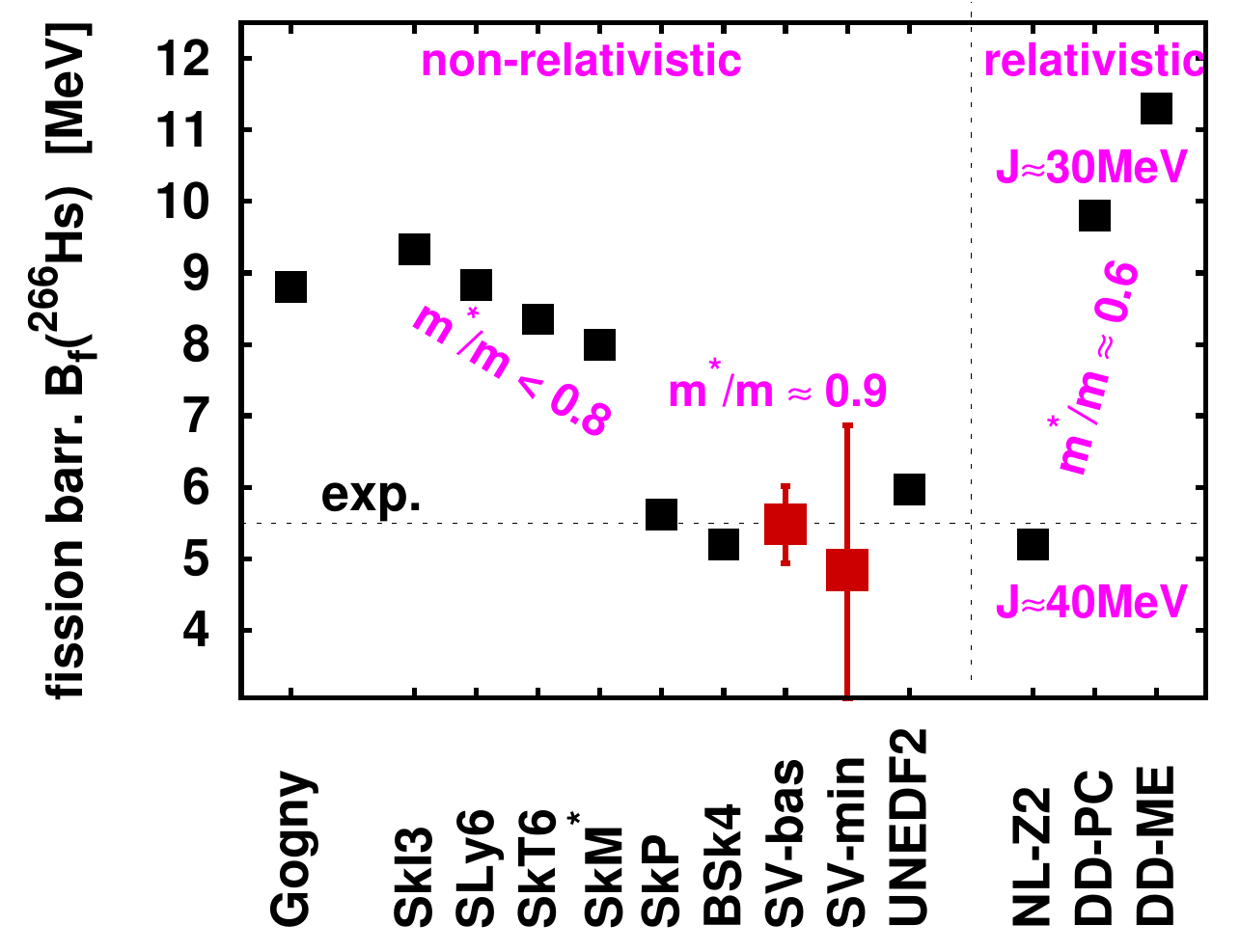}}
\caption{\label{fig:collect-barriers-266Hs}
Fission barrier $B_\mathrm{f}$ of $^{266}$Hs (Z=108) for a variety of
mean-field models and parametrizations. The values of two NMP
(symmetry energy $J$ and effective mass $m^*/m$) are indicated.
}
\end{figure}
A simple measure for fission stability is the fission barrier.  Figure
\ref{fig:collect-barriers-266Hs} shows the variety of $B_\mathrm{f}$
for $^{266}$Hs as predicted by different mean-field models, actually
the same selection for which we had shown the average quality in
figure \ref{quality-forces}. The span of $B_\mathrm{f}$ values is
large. But recall that the selection covers different generations of
parametrizations and that the calibration of models has much improved
over the years. The figure indicates two crucial NMP.
Parametrizations with $m^*/m\approx 0.9$ and $J\approx 30$ MeV deliver
best $B_f$.  Lower $m^*/m$ enhance shell effects, thus delivering
larger shell-correction energies at minimum and consequently larger
barriers.  There occurs a curious coincidence for the RMF
parametrization NL-Z2.  The effective mass is extremely low, but J is
extremely high. Both unusual NMP together happen to produce a correct
$B_\mathrm{f}$.

The error bars on the $B_\mathrm{f}$ for SV-bas and SV-min in figure
\ref{fig:collect-barriers-266Hs} show extrapolation uncertainties
obtained by statistical analysis (section \ref{sec:fit}). Both are
smaller than the span of results from the selection of
parametrizations. This is due to the fact that the large pool of data
in the recent fits produces better confined model parameters and, in
turn, smaller uncertainties in predicted observables. The large
difference in uncertainties between SV-bas and SV-min is explained by
the different fit data. While SV-min is fitted to ground-state data of
finite nuclei only, SV-bas fixes additionally the four NMP $K$, $m^*/m$,
$J$, and $\kappa_\mathrm{TRK}$ (which is equivalent to fit giant
resonances and polarizability in $^{208}$Pb \cite{Klu09a}). Putting
more information into SV-bas naturally reduces uncertainties. 
Interesting is the remarkable reduction as compared to SV-min. This
indicates that these four NMP together have a large impact on fission
barriers. This will be disentangled in more detail in the following
subsections.

\subsection{Impact of model parameters - trend analysis}
\label{sec:trend}

A simple and instructive way to explore the impact of model parameters
is trend analysis.  To this end, one defines a base parametrization,
chooses one particular model parameter, and varies it systematically
around the base point. Here we take as base point SV-bas \cite{Klu09a}
which was fitted where four NMP, namely $K$, $m/m$, $J$, and
$\kappa_\mathrm{TRK}$, had been kept fixed during the $\chi^2$ fit.
Then one takes one of these frozen NMP, sets it to a slightly
different value, and refits all other model parameters. This is done
for a couple of different values thus delivering a set of
parametrizations with systematically varied DFT parameter. Then one
can draw other observables as functions of this parameter which
eventually yields an impression of its impact on this observable.

\begin{figure}
\centerline{\includegraphics[width=\linewidth]{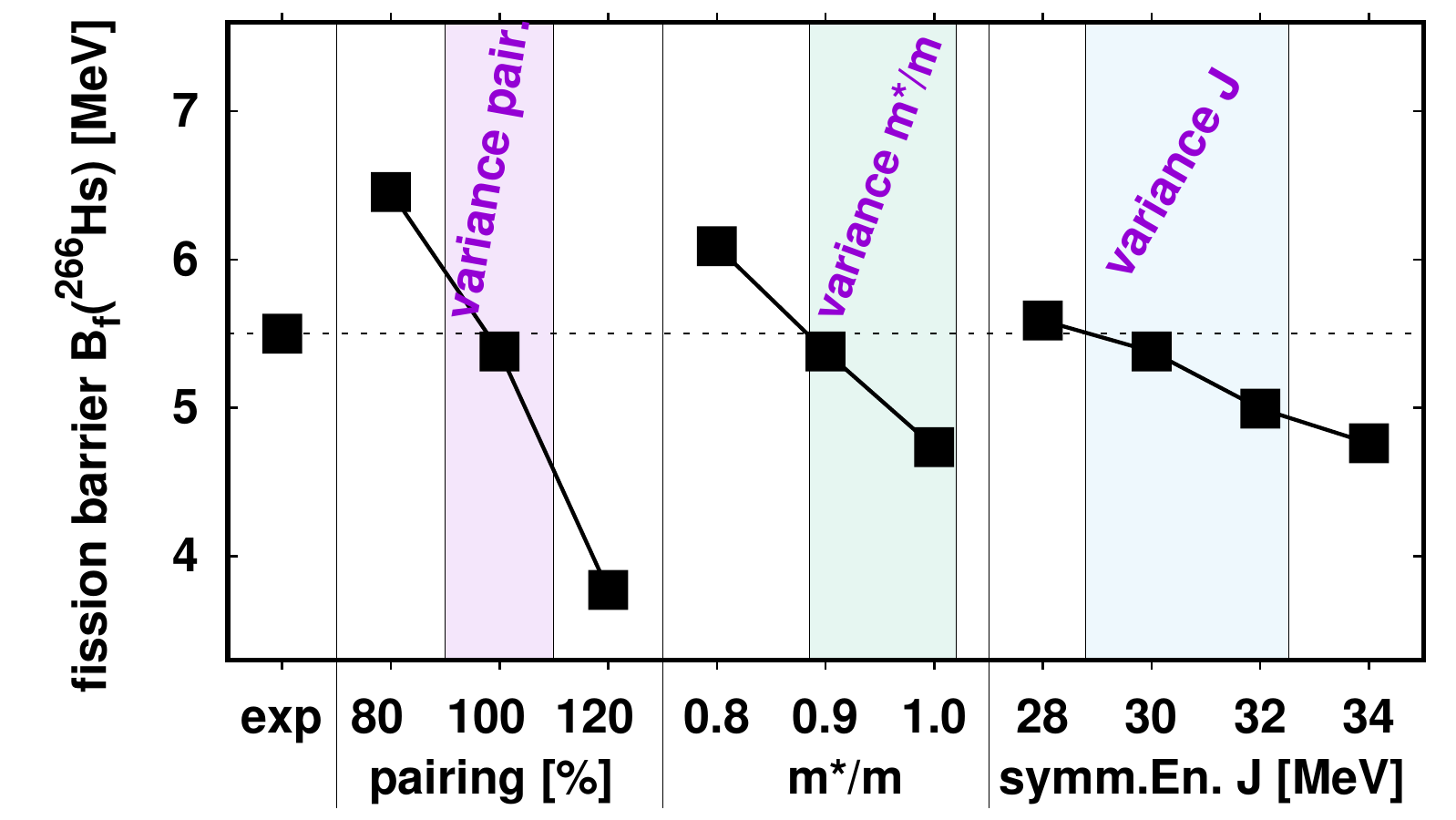}}
\caption{\label{fig:vary-known-she-TAN-short} Fission barrier $B_\mathrm{f}$ of
  $^{266}$Hs (Z=108) for three sets of SHF parametrizations with
  systematically varied properties, according to \cite{Klu09a}. The
  shaded areas indicate the uncertainty of each parameter.
}
\end{figure}
An example of this strategy is shown in figure
\ref{fig:vary-known-she-TAN-short} for the fission barrier
$B_\mathrm{f}$ of $^{266}$Hs as function of the three SHF parameters,
pairing strength, effective mass $m^*/m$, and symmetry energy $J$
which all were found to have visible effect on $B_\mathrm{f}$. Each
one of these three SHF parameters has an uncertainty according to the
rules of statistical analysis (see section \ref{sec:fit}) which is
indicated by a shaded area in the plot. Each parameter produces
significant changes of $B_\mathrm{f}$. It is interesting to note that
all three trends are very close to linear which confirms the expansion
about the optional parametrization usually employed in statistical
analysis.  The amount of variation of $B_f$ within the uncertainty
band indicates the strength of the correlation of $B_f$ with that
parameter and its contribution to its total uncertainty.  All three
parameters shown in the figure have about comparable and large impact.
The effect of pairing strength is plausible because pairing tends to
wash out the shell fluctuations and the fission barrier is produced by
the fluctuations of shell-correction energy with changing deformation.
Effective mass, again, is a quantity strongly related to shell
structure and influences $B_\mathrm{f}$ via shell-correction
energy. The symmetry energy $J$ is a true bulk property with ignorable
effect on shell structure. Its influence on $B_f$ is mediated
through its impact on neutron skin \cite{Naz10a,Naz14a} which, in
turn, plays a role in neck formation.

\subsection{Correlation  analysis}

\begin{figure}
\centerline{\includegraphics[width=\linewidth]{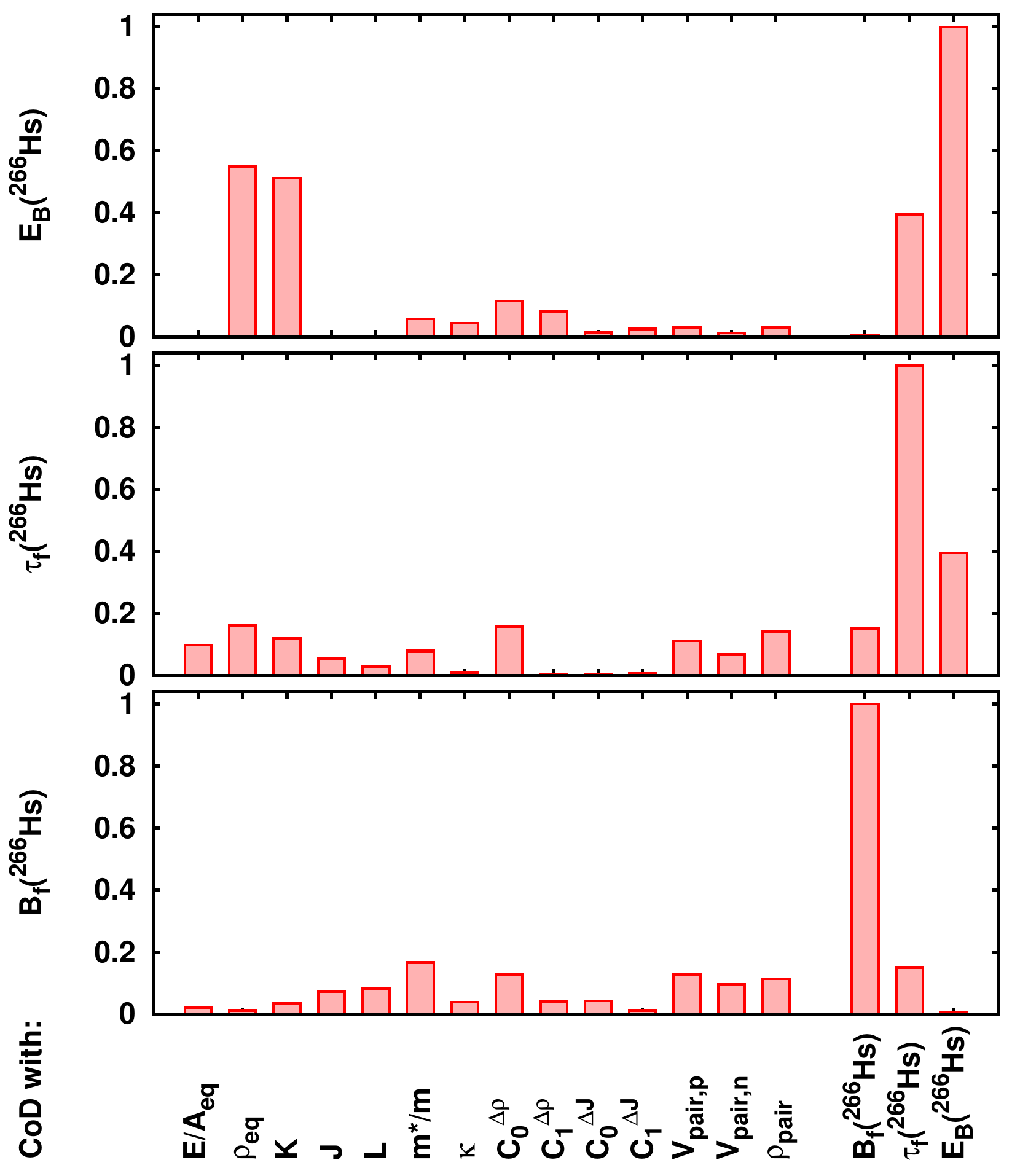}}
\caption{\label{fig:correl-fiss266Hs} Correlation of binding energy
  $E_B$, fission barrier $B_\mathrm{f}$, and fission halflife
  $\tau_\mathrm{f}$ of $^{266}$Hs with the 14 model parameters. The
  volume parameters are expressed in terms of NMP.  Correlations are
  quantified in terms of CoD.  }
\end{figure}
Trend analysis as exemplified in figure
\ref{fig:vary-known-she-TAN-short} is instructive. But it grows
cumbersome if more influences are to be compared. Correlation analysis
in terms of the alignment $r_{AB}$ or the CoD $r_{AB}^2$, see section
\ref{sec:fit}, provides a compact measure which allows to overview
many ingredients at once. Figure \ref{fig:correl-fiss266Hs} shows the
CoD between the SHF parameters and three observables in $^{266}$Hs,
binding energy $E_B$, fission lifetime $\tau_\mathrm{f}$, and fission
barrier $B_\mathrm{f}$.  \PGR{All three observables lack a prominent,
  strong correlation. They rather spread their influences over many
  parameters.  This holds, in particular, for the two fission
  observables $B_\mathrm{f}$ and $\tau_\mathrm{f}$.  For a proper
  interpretation, one has to keep in mind that not all SHF parameters
  are statistically independent from each other and that they have
  different degrees of determination.  The isoscalar, static bulk
  parameters $E/A_\mathrm{eq}$, $\rho_\mathrm{eq}$, $K$ and the
  isoscalar surface parameter $C_0^{\Delta\rho}$ are tightly fixed by
  the ground state data. They have too little variance to be used as
  lever for changing fission properties.  Among the parameters which
  leave more freedom, one has to take care of strong correlations
  (more than 90\% correlated): between the isovector properties $J$
  and $L$ and between the three pairing parameters. With these
  precautions, we see three somewhat more relevant influencers within
  the many SHF parameters, namely $J$, $m^*/m$, and pairing, exactly
  those three whose explicit variation was shown in figure
  \ref{fig:vary-known-she-TAN-short}.}

\PGR{The lack of prominent correlations for fission observables is
  surely a message. But one would still like to learn more about the
  impact of the model. This can be achieved by considering multiple
  correlation coefficients (MCC).}
\begin{figure}
\centerline{\includegraphics[width=\linewidth]{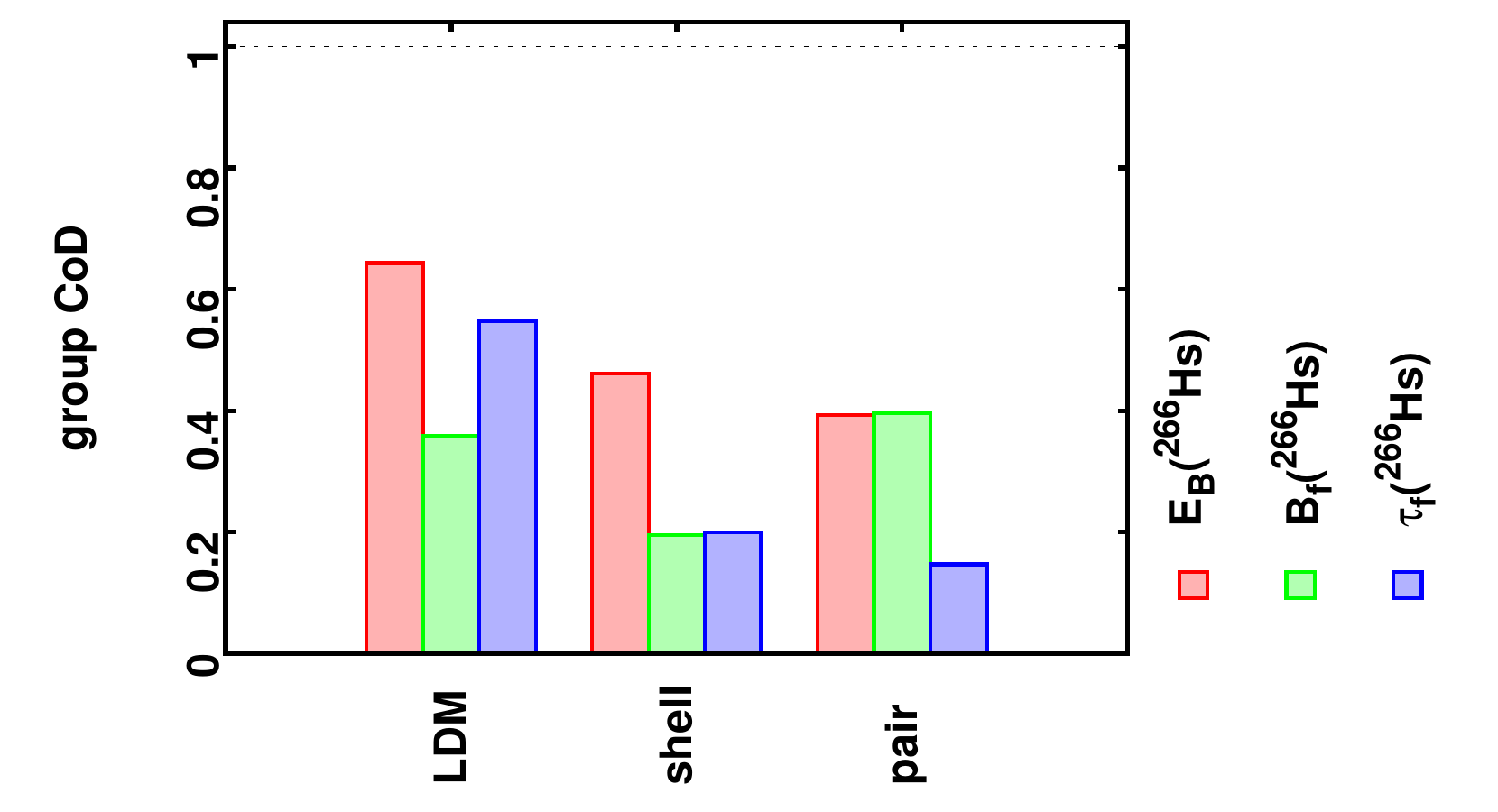}}
\caption{\label{fig:correl-group-fiss266Hs} Multiple correlation
  coefficient (MCC) of binding energy, fission barrier, as well as fission
  halflife of $^{266}$Hs with groups of model parameters, coined
  ``group CoD''.  The groups are: LDM = $E/A_\mathrm{eq}$,
  $\rho_\mathrm{eq}$, $K$, $J$, $L$, $C_0^{\nabla\rho}$, and
  $C_1^{\nabla\rho}$; shell = $m/m$, $\kappa_\mathrm{TRK}$,
  $C_0^{\nabla J}$, $C_1^{\nabla J}$; pair = $V_\mathrm{pair,p}$,
  $V_\mathrm{pair,n}$, $\rho_\mathrm{pair}$.  }
\end{figure}
For example, $V_\mathrm{pair,p}$, $V_\mathrm{pair,n}$, and
$\rho_\mathrm{pair}$ together embrace a group of parameters related to
pairing. One then can ask what is the impact of pairing as a whole
(and not only of one single pairing parameter) on fission. This is
quantified by the MCC
$R_{\mathbf{G}A}^2$ of an observable $A$ with groups of parameters
$\mathbf{G}=(p_{i_1},..,p_{i_G})$ \cite{Allison}.  Values of $R^2$,
again, range from 0 to 1, where 0 implies, that those quantities are
completely uncorrelated, 1 denotes that the group $\mathbf{G}$
determines the observable $A$ completely. An $R^2$ of, say, $0.30$
means that 30\% of the variance in $A$ is predictable from
$\mathbf{G}$.  For a group containing all model parameters, an
observable is completely determined, hence $R^2=1$. Figure
\ref{fig:correl-group-fiss266Hs} shows MCC of fission properties in
$^{266}$Hs with three groups of parameters as indicated.  The
liquid-drop-model (LDM) embraces the static NMP together with
isoscalar and isovector surface parameters. It represents the general
trends of nuclear bulk properties averaging through shell fluctuations
\cite{Mye82aR}.  The ``shell'' group collects the four parameters
having direct impact on level structure, particularly level density
and the ``pair'' group is obviously related to the three pairing
parameters in the model.  Although the MCC are considerably larger
than the single CoD in figure \ref{fig:correl-fiss266Hs}, the basic
feature remains, namely that the influences are distributed, here over
the three groups under consideration. This illustrates once again that
properties of SHE emerge from a subtle interplay of all ingredients of
the model. A safe prediction requires careful modeling and counter
checks in the SHE region.

\subsection{Trends with nucleon number}

\begin{figure}
\centerline{\includegraphics[width=0.8\linewidth]{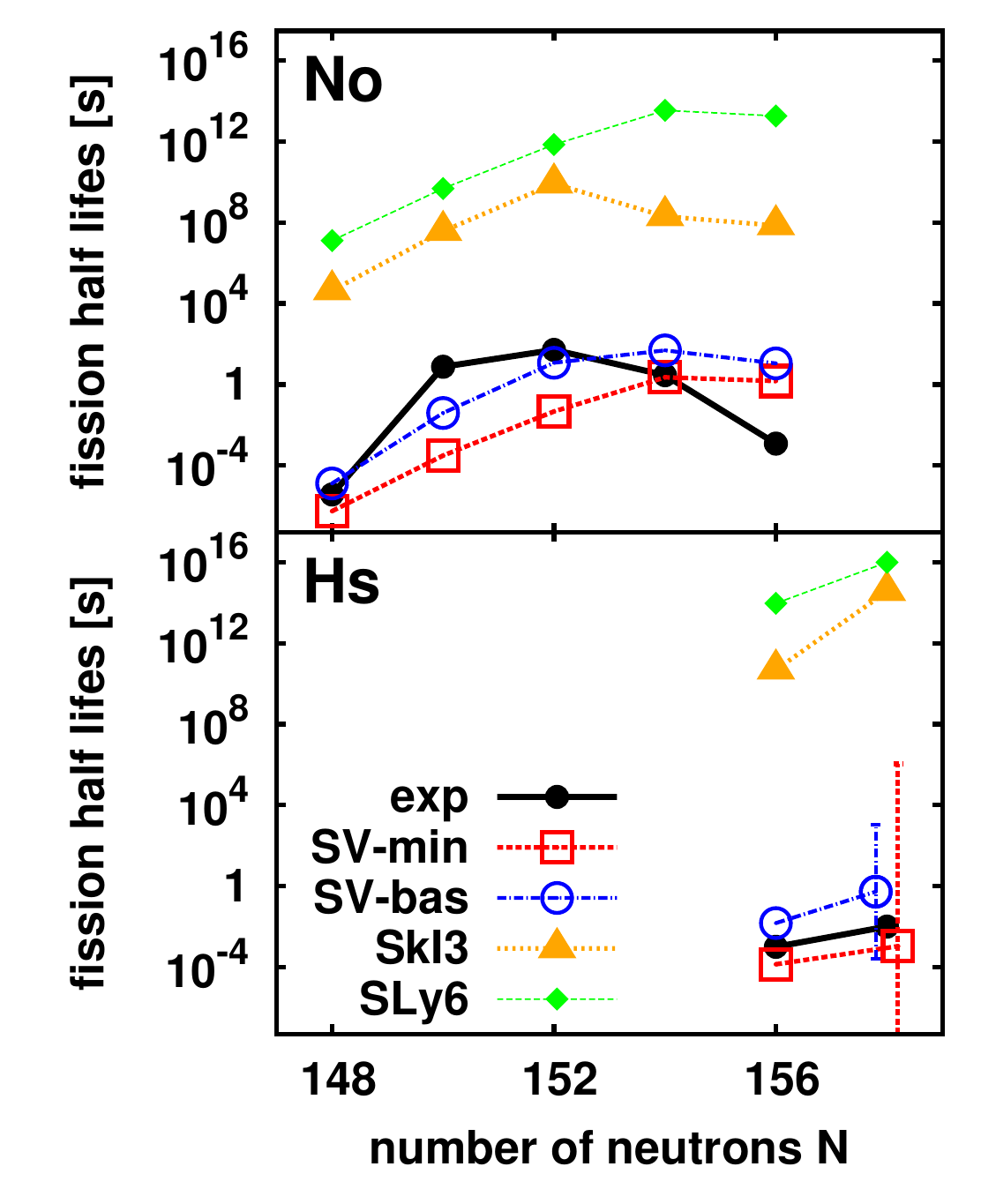}}
\caption{\label{fig:halflifes-No-Hs} Fission halflifes $\tau_f$ along
  isotopic chains of two transactinides, No and Hs. Results from four
  different SHF parametrizations are compared with experimental data
  (for complete references to experimental fission data see
  \cite{Erl12b}). Approximate extrapolation errors are indicated for
  $^{266}$Hs. }
\end{figure}
Having explored the impact of model parameters on fission properties,
we have finally a quick glance at the system dependence and do this
for the fission lifetime $\tau_f$. Figure \ref{fig:halflifes-No-Hs}
shows the isotopic trends of lifetimes for two SHE, the transactinides
No and Hs for which experimental data exist. The data are compared
with results from four SHF parametrizations out of the selection of
parametrizations in figure \ref{fig:collect-barriers-266Hs} where two
of them (SV-min and SV-bas) delivered barriers close to data and the
other two had too high barriers. The consequence on lifetimes is
dramatic as we see in the figure \ref{fig:halflifes-No-Hs} here.  SLy6
and SkI3 which overestimate barriers in $^{266}$Hs by 3-4 MeV produce
lifetimes more than 12 orders of magnitude too large. SV-min and
SV-bas whose barriers reach data within less than 1 MeV reproduce
lifetimes within 2 orders of magnitude. An estimate of the
extrapolation uncertainty is shown for $^{266}$Hs. It amounts to hefty
9 orders of magnitude for the unrestricted fit SV-min. Only with
additional information on response properties built into SV-bas, we
come down to about 3-4 orders of magnitude uncertainty. This is a
humble figure of what theory can promise at its best.  The example
thus demonstrates once more the extreme sensitivity of fission
lifetimes. A reproduction of data within four orders of magnitude can
be considered as success.  This goal can be reached with careful
calibration of the mean-field model. With the latest generation of SHF
parametrizations we dispose now of a reliable description of fission
properties of SHE. This can then be used for large-scale surveys in
the whole landscape of SHE establishing systematics in the available
region and exploring also extremes as, e.g., SHE in the $r$-process
region as has been done, e.g., in \cite{Erl12b,Pet12a}.

\section{Conclusions and outlook}

We have reviewed the theoretical prediction of fission properties of
super-heavy elements (SHE) by self-consistent mean-field models.  For
the latter, we considered particularly the widely used
Skyrme-Hartree-Fock (SHF) model. The theoretical tools, namely the
energy-density functional of SHF, its calibration by least-squares
fits with subsequent statistical analysis, and the computation of
fission lifetimes, are briefly explained. Already at the formal level,
one sees that fission calculations are very involved and the final
result emerges from a subtle interplay of several ingredients. 
Statistical analysis is the tool of choice to disentangle the impact
of the various effects and to get an idea of the predictive value of
mean-field models.

We have given a summary overview of fission properties in the
landscape of SHE in comparison to competing decay channels ($\alpha$-
and $\beta$-decay, neutron emission). Using a carefully chosen SHF
parametrization, the known properties of SHE are well reproduced which
gives some confidence in the extrapolation deep into the $r$-process
region.  Most of the discussion was concerned with analyzing the
reliability of SHF for computing properties of SHE. Taking published
SHF parametrizations collected over decades of development, yields a
disquietingly large spread of predictions. But it shows also that
there are many parametrizations which perform very well and this can
be related to response properties as, e.g., effective mass or symmetry
energy. The impact of these and other properties is worked out by
trend and correlation analysis. It is found, quite as expected, that
the sensitivity of fission properties is spread over a great multitude
of model parameters without clear preference. More can be said when
grouping the model parameters into sensible blocks. We have defined
three such blocks, a group of liquid-drop model (LDM) properties
(static bulk properties without shell effects), a pairing group, and a
group determining shell fluctuations. It turns out that the LDM group
has largest impact, but not an exclusive one. With comparable weight
contributes the pairing group to fission lifetimes and the group of
shell effects to the barrier. Finally, we have checked possible trends
with system size and charge within the landscape of transactinides.
There seems to be no trend in agreement with data. Well fitting
parametrizations do that all over this landscape and those who fail
fail throughout.  Still all results together show that predictions of
fission properties have to be taken with care.
When extending them to new regions, counter checking
with one or two reference points is highly
recommended.

\section*{Acknowledgments}
First of all, my thank goes to Walter Greiner who inspired and
accompanied the studies on self-consistent mean-field models in
connection with super-heavy elements. There were many contributors to
this research over the years which we cannot list here in detail. But
at least two amongst them deserve mentioning, namely J.A. Maruhn and
W. Nazarewicz. Their active involvement together with their groups was
the necessary condition to work out the large body of work on SHE and
nuclear energy density functionals.

\bibliographystyle{epj}
\bibliography{FoS}

\begin{thebibliography}{66}

\bibitem{Hyd57aR}
E.K. Hyde, G.T. Seaborg, in \emph{Handbuch der Physik, Vol. XLII} (Springer,
  Berlin, 1957), p. 205

\bibitem{Nix72aR}
J.R. Nix, Annual Review of Nuclear Science \textbf{22}, 66 (1972)

\bibitem{Hof00aER}
S.~Hofmann, G.~M{\"u}nzenberg, Rev. Mod. Phys. \textbf{72}, 733 (2000)

\bibitem{HofS01}
S.~Hofmann, F.~He{\ss}berger, D.~Ackermann, S.~Antalic, P.~Cagarda, S.~Ćwiok,
  B.~Kindler, J.~Kojouharova, B.~Lommel, R.~Mann et~al., Eur. Phys. J. A
  \textbf{10}, 5 (2001), ISSN 1434-6001

\bibitem{Oga04}
Y.T. Oganessian, V.K. Utyonkov, Y.V. Lobanov, F.S. Abdullin, A.N. Polyakov,
  I.V. Shirokovsky, Y.S. Tsyganov, G.G. Gulbekian, S.L. Bogomolov, B.N. Gikal
  et~al., Phys. Rev. C \textbf{70}(6), 064609 (2004)

\bibitem{Oga06}
Y.T. Oganessian, V.K. Utyonkov, Y.V. Lobanov, F.S. Abdullin, A.N. Polyakov,
  R.N. Sagaidak, I.V. Shirokovsky, Y.S. Tsyganov, A.A. Voinov, G.G. Gulbekian
  et~al., Phys. Rev. C \textbf{74}(4), 044602 (2006)

\bibitem{Greg06}
K.E. Gregorich, J.M. Gates, C.E. D\"ullmann, R.~Sudowe, S.L. Nelson, M.A.
  Garcia, I.~Dragojevi\ifmmode~\acute{c}\else \'{c}\fi{}, C.M. Folden~III, S.H.
  Neumann, D.C. Hoffman et~al., Phys. Rev. C \textbf{74}(4), 044611 (2006)

\bibitem{Dvo08}
J.~Dvorak, W.~Br\"uchle, M.~Chelnokov, C.E. D\"ullmann, Z.~Dvorakova,
  K.~Eberhardt, E.~J\"ager, R.~Kr\"ucken, A.~Kuznetsov, Y.~Nagame et~al., Phys.
  Rev. Lett. \textbf{100}(13), 132503 (2008)

\bibitem{Martinez-Pinedo.Mocelj.ea:2007}
G.~Mart{\'i}nez-Pinedo, D.~Mocelj, N.~Zinner, A.~Kelić, K.~Langanke, I.~Panov,
  B.~Pfeiffer, T.~Rauscher, K.H. Schmidt, F.K. Thielemann, Prog. Part. Nucl.
  Phys. \textbf{59}(1), 199 (2007)

\bibitem{Arn07}
M.~Arnould, S.~Goriely, K.~Takahashi, Physics Reports \textbf{450}(4-6), 97
  (2007)

\bibitem{Sto07aR}
J.~Stone, P.G. Reinhard, Prog. Part. Nucl. Phys. \textbf{58}, 587 (2007)

\bibitem{Mos68a}
U.~Mosel, W.~Greiner, Z. Phys. \textbf{217}, 256 (1968)

\bibitem{Mos69a}
U.~Mosel, W.~Greiner, Z. Phys. \textbf{222}, 261 (1969)

\bibitem{Rin96aR}
P.~Ring, Prog. Part. Nucl. Phys. \textbf{37}, 193 (1996)

\bibitem{Ben03aR}
M.~Bender, P.H. Heenen, P.G. Reinhard, Rev. Mod. Phys. \textbf{75}, 121 (2003)

\bibitem{Vre05aR}
D.~Vretenar, A.~Afanasjev, G.~Lalazissis, P.~Ring, Phys. Rep. \textbf{409}, 101
  (2005)

\bibitem{Erl11a}
J.~Erler, P.~Kl\"upfel, P.G. Reinhard, J. Phys. G \textbf{38}, 033101 (2011)

\bibitem{Bra72aR}
M.~Brack, J.~Damg{\aa}rd, A.S. Jensen, H.C. Pauli, V.M. Strutinsky, C.Y. Wong,
  Rev. Mod. Phys. \textbf{44}, 320 (1972)

\bibitem{Ber01a}
J.F. Berger, L.~Bitaud, J.~Dechargé, M.~Girod, K.~Dietrich, Nuclear Physics A
  \textbf{685}(1-4), 1  (2001)

\bibitem{War06}
M.~Warda, J.~Egido, L.~Robledo, Phys. Scr. T \textbf{125}, 226 (2006)

\bibitem{Sta09a}
A.~Staszczak, A.~Baran, J.~Dobaczewski, .W. Nazarewicz, Phys. Rev. C
  \textbf{80}(1), 014309 (2009)

\bibitem{Sch09a}
N.~Schindzielorz, J.~Erler, P.~Kl\"upfel, P.G. Reinhard, G.~Hager,
  International Journal of Modern Physics E \textbf{18}(4), 773 (2009)

\bibitem{Erl12b}
J.~Erler, K.~Langanke, H.P. Loens, G.~Martinez-Pinedo, P.G. Reinhard, Phys.
  Rev. C \textbf{85}, 025802 (2012)

\bibitem{Bar15aR}
A.~Baran, M.~Kowal, P.G. Reinhard, L.~Robledo, A.~Staszczak, M.~Warda, Nucl.
  Phys. A \textbf{944}, 442 (2015)

\bibitem{Ben07a}
T.~Lesinski, M.~Bender, K.~Bennaceur, T.~Duguet, J.~Meyer, Phys. Rev. C
  \textbf{76}, 014312 (2007)

\bibitem{Rei75a}
P.G. Reinhard, Nucl. Phys. A \textbf{252}, 120 (1975)

\bibitem{Klu08a}
P.~Kl\"upfel, J.~Erler, P.G. Reinhard, J.A. Maruhn, Eur. Phys. J A \textbf{37},
  343 (2008)

\bibitem{Klu09a}
P.~Kl\"upfel, P.G. Reinhard, T.J. B\"urvenich, J.A. Maruhn, Phys.Rev. C
  \textbf{79}, 034310 (2009)

\bibitem{Rin80aB}
P.~Ring, P.~Schuck, \emph{The Nuclear Many-Body Problem} (Springer--Verl., New
  York, Heidelberg, Berlin, 1980)

\bibitem{Bon85a}
P.~Bonche, H.~Flocard, P.H. Heenen, S.J. Krieger, M.S. Weiss, Nucl. Phys. A
  \textbf{443}, 39 (1985)

\bibitem{Kri90a}
S.J. Krieger, P.~Bonche, H.~Flocard, P.~Quentin, M.S. Weiss, Nucl. Phys. A
  \textbf{517}, 275 (1990)

\bibitem{Ben00c}
M.~Bender, K.~Rutz, P.G. Reinhard, J.A. Maruhn, Eur. Phys. J. A \textbf{8}, 59
  (2000)

\bibitem{Bev69aB}
P.R. Bevington, D.K. Robinson, \emph{Data Reduction and Error Analysis for the
  Physical Sciences} (McGraw-Hill, 2003)

\bibitem{Fri86a}
J.~Friedrich, P.G. Reinhard, Phys. Rev. C \textbf{33}, 335 (1986)

\bibitem{Nik04a}
T.~Niksic, D.~Vretenar, G.~Lalazissis, P.~Ring, Phys. Rev. C \textbf{69},
  047301 (2004)

\bibitem{Nik08a}
T.~Niksic, D.~Vretenar, P.~Ring, Phys. Rev. C \textbf{78}, 034318 (2008)

\bibitem{Kor12a}
M.~Kortelainen, J.~McDonnell, W.~Nazarewicz, P.G. Reinhard, J.~Sarich,
  N.~Schunck, M.V. Stoitsov, S.M. Wild, Phys. Rev. C \textbf{85}, 024304 (2012)

\bibitem{Bro98a}
B.A. Brown, Phys. Rev. C \textbf{58}, 220 (1998)

\bibitem{Kor14a}
M.~Kortelainen, J.~McDonnell, W.~Nazarewicz, E.~Olsen, P.G. Reinhard,
  J.~Sarich, N.~Schunck, S.M. Wild, D.~Davesne, J.~Erler et~al., Phys. Rev. C
  \textbf{89}, 054314 (2014)

\bibitem{Hag08a}
G.~Hagen, T.~Papenbrock, D.J. Dean, M.~Hjorth-Jensen, Phys. Rev. Lett.
  \textbf{101}, 092502 (2008)

\bibitem{Nav09a}
P.~Navr\'atil, S.~Quaglioni, I.~Stetcu, B.R. Barrett, J. Phys. G \textbf{36},
  083101 (2009)

\bibitem{Gar16a}
R.F.G. Ruiz, M.L. Bissell, K.~Blaum, A.~Ekström, N.~Frömmgen, G.~Hagen,
  M.~Hammen, K.~Hebeler, J.D. Holt, G.R. Jansen et~al., Nature Physics
  \textbf{12}, 594 (2016)

\bibitem{Bei75a}
M.~Beiner, H.~Flocard, {Nguyen Van Giai}, P.~Quentin, Nucl. Phys. A
  \textbf{238}, 29 (1975)

\bibitem{Bar82a}
J.~Bartel, P.~Quentin, M.~Brack, C.~Guet, H.B. H{\aa}kansson, Nucl. Phys. A
  \textbf{386}, 79 (1982)

\bibitem{Dob84a}
J.~Dobaczewski, H.~Flocard, J.~Treiner, Nucl. Phys. A \textbf{422}, 103 (1984)

\bibitem{Ton84a}
F.~Tondeur, M.~Brack, M.~Farine, J.M. Pearson, Nucl. Phys. \textbf{A420}, 297
  (1984)

\bibitem{Rei95a}
P.G. Reinhard, H.~Flocard, Nucl. Phys. A \textbf{584}, 467 (1995)

\bibitem{Cha98a}
E.~Chabanat, P.~Bonche, P.~Haensel, J.~Meyer, R.~Schaeffer, Nucl. Phys. A
  \textbf{635}, 231 (1998), {N}ucl. {P}hys. \textbf{A643}, 441(E)

\bibitem{Gor03a}
S.~Goriely, M.~Samyn, M.~Bender, J.M. Pearson, Phys. Rev. C \textbf{68}, 054325
  (2003)

\bibitem{Gor13a}
S.~Goriely, N.~Chamel, J.M. Pearson, Phys. Rev. C \textbf{88}, 061302(R) (2013)

\bibitem{Dob01b}
J.~Dobaczewski, W.~Nazarewicz, P.G. Reinhard, Nucl. Phys. \textbf{A693}, 361
  (2001)

\bibitem{Glantz}
S.A. Glantz, B.K. Slinker, T.B. Neilands, \emph{Primer of Applied Regression \&
  Analysis of Variance} (McGraw Hill, 1990)

\bibitem{Rut95a}
K.~Rutz, J.A. Maruhn, P.G. Reinhard, W.~Greiner, Nucl. Phys. \textbf{A590}, 680
  (1995)

\bibitem{Cus85a}
R.Y. Cusson, P.G. Reinhard, M.R. Strayer, J.A. Maruhn, W.~Greiner, Z. Phys. A
  \textbf{320}, 475 (1985)

\bibitem{Rei87aR}
P.G. Reinhard, K.~Goeke, Rep. Prog. Phys. \textbf{50}, 1 (1987)

\bibitem{Dob07a}
J.~Dobaczewski, M.~Stoitsov, W.~Nazarewicz, P.G. Reinhard, Phys. Rev. C
  \textbf{76}, 054315 (2007)

\bibitem{Rei78a}
P.G. Reinhard, Nucl. Phys. A \textbf{306}, 19 (1978)

\bibitem{Rei84b}
P.G. Reinhard, F.~Gr\"ummer, K.~Goeke, Z. Phys. \textbf{A317}, 339 (1984)

\bibitem{Vio66a}
V.~Viola, G.~Seaborg, Journal of Inorganic and Nuclear Chemistry \textbf{28},
  741 (1966)

\bibitem{Pot10a}
K.J. Pototzky, J.~Erler, P.G. Reinhard, V.O. Nesterenko, subm. Eur. Phys. A
  (2010)

\bibitem{Uma15a}
A.S. Umar, V.E. Oberacker, C.J. Horowitz, P.G. Reinhard, J.A. Maruhn, Phys.
  Rev. C \textbf{92}, 025808 (2015)

\bibitem{Naz10a}
P.G. Reinhard, W.~Nazarewicz, Phys. Rev. C \textbf{81}(5), 051303 (2010)

\bibitem{Naz14a}
W.~Nazarewicz, P.G. Reinhard, W.~Satu{\l}a, D.~Vretenar, Eur. Phys. J. A
  \textbf{50}, 20 (2014)

\bibitem{Allison}
P.D. Allison, \emph{Multiple Regression: A Primer} (Sage Publications, 1998)

\bibitem{Mye82aR}
W.D. Myers, W.J. Swiatecki, Annual Review of Nuclear and Particle Science
  \textbf{32}, 309 (1982)

\bibitem{Pet12a}
I.~Petermann, K.~Langanke, G.~Martinez-Pinedo, P.G.R. I.V.~Panov, F.K.
  Thielemann, Eur. Phys. J. A \textbf{48}, 122 (2012)

\end{thebibliography}

\end{document}